\documentstyle[12pt]{article}
\begin{document}
{\pagestyle{empty}
\rightline{NWU-11/95}
\rightline{Sept. 1995}
\rightline{~~~~~~~~~~}
\vskip 1cm
\centerline{\Large \bf Classical and Quantum Solutions and the Problem of
Time }
\centerline{\Large \bf in $R^2$ Cosmology}
\vskip 1cm

\centerline{Masakatsu Kenmoku
\footnote{E-mail address:kenmoku@nara-wu.ac.jp}
and
Kaori Otsuki
\footnote{E-mail address: sp19254@nara-wu.ac.jp}
}
\centerline{{\it Department of Physics} }
\centerline{{\it Nara Women's University, Nara 630, Japan} }

\vskip 0.6cm

\centerline{Kazuyasu Shigemoto
\footnote{E-mail address:shigemot@tezukayama-u.ac.jp}
and
Kunihiko Uehara
\footnote{E-mail address:uehara@tezukayama-u.ac.jp}
}
\centerline {{\it Department of Physics}}
\centerline {{\it Tezukayama University, Nara 631, Japan }}

\vskip 1cm

\centerline{\bf Abstract}
\vskip 0.2in

We have studied various classical solutions in $R^2$ cosmology.
Especially we have obtained general classical solutions in
pure $R^2$\ cosmology.
Even in the quantum theory, we can solve the Wheeler-DeWitt equation
in pure $R^2$\ cosmology exactly. Comparing these classical and quantum
solutions in $R^2$\ cosmology, we have studied the problem of time in
general relativity.

\vskip 0.4cm
\noindent
\hfil
\vfill
\newpage}

\vspace{1cm}

\section{Introduction}

\vspace{0.5cm}

\indent

Recently quantum theoretical treatment of gravity has been paid
attention in various topics such as the time development of quite
early universe\cite{Hawking,Vilenkin}, exact treatment of Hawking
radiation\cite{Tomimatsu}\ and the problem of time in
cosmology\cite{Banks}-\cite{Ohkuwa}.

On the other hand, attempts to generalize Einstein's gravitation theory
have a long history in various purposes.
The generalized Einstein theory with additional $R^2$ term was
introduced to regularize ultraviolet divergences in the Einstein
theory\cite{Utiyama}.  It was applied to cosmology to obtain the
bounce universe to avoid the singularity at the
creation of the universe\cite{Nariai}.
The structure and the properties of the higher derivative gravitation
theory were further elaborated in subsequent works\cite{Higher}.

Other people have tried to generalize Einstein theory to modify Newtonian
force laws from the framework of general
relativity\cite{Brans,Fujii}.
Since there is no established direct observation
of dark matter, there are many attempts to explain
rotation curves without dark matter by modifying the Newtonian
force empirically\cite{Sanders} or from the framework of the
general relativity\cite{Ours}. Whereas, on the application of the generalized
Einstein theory to cosmology, many people are interested in $R^2$ cosmology
as it may give a promising model to explain the inflation in early universe
without using the fictitious scalar field\cite{Suen,Mijic}.
The semiclassical solution of higher-order cosmology gives the de Sitter
universe\cite{Starobinsky}, but there arises instability in the quantum
treatment\cite{Schmidt}.

In the previous paper, we attempted to explain not
only the flat rotation curves of spiral galaxies but also
the large-scale structure of the universe, starting
from a simple model with adding $R^2$\ term to the
Einstein action\cite{Ours}.
For the generalized Einstein action, we have the quite
stringent observational constraint that
the coefficient $\gamma$ of the Robertson expansion\cite{Weinberg}
must be quite near $1$ from the classical tests of
general relativity by the radar echo delay experiment.  However,
in the previous paper, we constructed the model which gives
$\gamma=1$ at solar scale and gives
Sander's type gravitational potential at galaxy scale\cite{Ours2}.

In series of our papers, we applied our $R^2$ involved gravitational
theory to {\it 1)}~ solar scale, {\it 2)}~ galaxy scale, {\it 3)}~
large-scale of the universe, such as "Great Wall", in classical level.
Then in this paper, we will apply this $R^2$ involved gravitational theory
to cosmology. First we will find various classical solutions
in $R^2$ cosmology.  Especially we will give general classical
solutions in pure $R^2$\ cosmology.  Next we will consider the
quantum theory of pure $R^2$\ cosmology, and will give exact solutions
of Wheeler-DeWitt equation.
Comparing these classical and quantum solutions in $R^2$\ cosmology,
we will study the problem of time in the general relativity according to
the de Broglie-Bohm interpretation\cite{dBB}.

\vspace{1cm}


\section{Classical Solutions of $R^2$ Cosmology}

\vspace{0.5cm}

\indent

We use Weinberg notation\cite{Weinberg}\ and consider the following
generalized action:

\begin{eqnarray}
I&~=\displaystyle{ \int} d^4x \sqrt{-g} \left[
-\displaystyle{ \frac{1}{16 \pi G}}(R + \alpha R^2 )
+{\cal L_{\rm matter}} \right]~, \label{e1}
\end{eqnarray}

\noindent
where $G$\ is the gravitational constant, $R$\ is the scalar curvature
and $\alpha$ is a constant.

In order to study classical and quantum solutions in $R^2$\ cosmology,
we consider the homogeneous and the isotropic space-time with the metric

\begin{equation}
ds^2=-b(t)^2 {dt}^2+a(t)^2\left[ {dr}^2
+r^2 ({d\theta}^2 +\sin{\theta}^2 {d\varphi}^2) \right]~,  \label{e2}
\end{equation}


\noindent
where we put the curvature term to be zero.

Using this metric, we obtain the expression of the scalar curvature

\begin{eqnarray}
R&=&-6 \left( { \ddot{a} \over {a b^2} }+{\dot{a}^2 \over {a^2 b^2}}
-{{\dot{a} \dot{b}} \over {a b^3}}  \right).~ \label{e3}
\end{eqnarray}

After partial integration, the original action can be transformed
into the form

\begin{eqnarray}
I=\displaystyle{ \int} d^3 x \displaystyle{\int} dt
 \left[
     - {3 \over {8\pi G}}
     \left\{
       {{a \dot{a}^2}  \over  b}
       +6\alpha
        \left( { {\ddot{a}^2 a}  \over  b^3}
             -{ {2\ddot{a}\dot{a}\dot{b} a}  \over  b^4}
             +{ {\dot{a}^2\dot{b}^2 a}  \over  b^5}
             +{ {\dot{a}^4}  \over  {a b^3}}
         \right)
    \right\}
    + a^3 b {\cal L_{\rm matter}}
 \right].
\label{e4}
\end{eqnarray}

For perfect fluid approximation of matter, equations of motion
to be solved become

\begin{eqnarray}
&& H^2-6\alpha\left\{ 2 \ddot{H} H -\dot{H}^2
 +6 \dot{H} H^2\right\}={{8 \pi G} \over 3} \rho , \label{e7} \\
&& \dot{\rho}+3 H (\rho+P)=0, \label{e8}
\end{eqnarray}

\noindent
where we put $b=1$ after variation.  We use notation $H=\dot{a}/a$, and
denote $\rho$ and $P$ as density and pressure of
perfect fluid respectively.

\vspace{1cm}


\subsection{Classical solution with perfect fluid in pure $R^2$ cosmology}

\vspace{0.5cm}

\indent

We put $P=\gamma \rho$ in this case, then equations to be solved become
in the form:

\begin{eqnarray}
&& -6\alpha\left\{ 2 \ddot{H} H -\dot{H}^2
 +6 \dot{H} H^2\right\}={{8 \pi G} \over 3} \rho , \label{e9} \\
&& \dot{\rho}+3 H (1+\gamma)\rho=0. \label{e10}
\end{eqnarray}

\noindent
{}From Eq.(\ref{e10}), we obtain $\rho a^{3(1+\gamma)}
=\rho_0 {a_0}^{3(1+\gamma)}=({\rm const.})$.
Substituting this relation into Eq.({\ref{e9}), we obtain the
following power type solution of the universe:

\begin{eqnarray}
a(t)&=& k_1 t^{4 \over {3(1+\gamma)}} , \label{e11} \\
{\rm with \quad} k_1&=&a_0 \left({{\pi G \rho_0 (1+\gamma)^3} \over
{4 \alpha (5-3\gamma)}}\right)^{1 \over {3(1+\gamma)}} , \nonumber
\end{eqnarray}
\noindent
where we use the initial condition $a(0)=0$.


\vspace{1cm}


\subsection{Classical solution without matter in $R + \alpha R^2$
cosmology}

{\bf a) { Case of $\alpha<0$}}

\vspace{0.5cm}

In this case, equation to be solved becomes in the form\cite{Suen,Mijic}:

\begin{eqnarray}
H^2-6\alpha\left\{ 2 \ddot{H} H -\dot{H}^2
 +6 \dot{H} H^2\right\}=0. \label{e12}
\end{eqnarray}

\noindent
When we denote ${\tilde \alpha}=-\alpha(>0)$ and
change variables into forms

\begin{eqnarray}
H(t)&=& {\sqrt{6} \over {36 \sqrt{\tilde{\alpha}}}}~ f(\xi)^2,  \nonumber \\
t&=&\sqrt{24 \tilde{\alpha}}~ \xi,  \label{e13}
\end{eqnarray}

\noindent
we obtain following equation:

\begin{eqnarray}
{{d^2 f} \over {d \xi^2}}+f^2 {df \over {d \xi}}+f=0,  \label{e14}
\end{eqnarray}

\noindent
except the uninteresting trivial case $f=0$~ (static universe).

Though we cannot find exact solution for Eq.(\ref{e14}), we can solve that
equation perturbatively.
First, treating the third term in Eq.(\ref{e14}) as the perturbation,
we obtain the following  power series solution:

\begin{eqnarray}
f(\xi)=\sqrt{3 \over {2 \xi}} \left[ 1-{2 \over 3} \xi^2
-{2 \over 25} \xi^4 +{4 \over 4725} \xi^6 +{23186 \over 4606878} \xi^8
+{6428 \over 4558125} \xi^{10} +\cdots \right]~ ,
\label{e15}
\end{eqnarray}

\noindent
where we assumed that singularity occures at $\xi=0$.

Second, treating the second term in Eq.(\ref{e14}) as the perturbation,
we obtain the following oscillating solution by taking the initial
condition $f(0)=0$:

\begin{eqnarray}
f(\xi)=\sin{\xi} +\left[{1 \over 32} (\cos{\xi}-\cos{3 \xi})-{\xi \over 8}
\sin{\xi}\right]+ \cdots . \label{e16}
\end{eqnarray}

Combining these perturbative solutions, we know the asymptotic
behavior $f(\xi) \sim \sin(\xi-\xi_0) / \sqrt{\xi}$
as $\xi \rightarrow \infty$.

Here we must notice that the above solution does not approach
the solution of equation
 $ H^2=0 $, which comes from
Eq.(\ref{e12}) by putting $\tilde{\alpha}=0$\ from the beginning, even
in the limit $\tilde{\alpha} \rightarrow 0$.
This is because ${\tilde \alpha}$ is
the coefficient of the highest derivative term. Then the limit
$\tilde{\alpha} \rightarrow 0$ give the quite chaotic behavior, which
is quite similar to the turbulent limit in the Navier-Stokes equation.
Actually, $H(t)$ has the following $t$~ dependence:

\begin{eqnarray}
H(t)={\sqrt{6} \over {36 \sqrt{\tilde{\alpha}}}}~
f^2({t \over {\sqrt{24 \tilde{\alpha}}}}). \label{e17}
\end{eqnarray}

\noindent
In the limit $\tilde{\alpha} \rightarrow 0$\ with finite $t$,
$H(t)$ oscillates quite rapidly and the oscillation amplitude
diverges, and finally becomes chaotic.

\vspace{0.7cm}

{\bf b) { Case of $\alpha>0$}}

\vspace{0.5cm}

We define ${\tilde \alpha}=\alpha(>0)$\ in this case and use
the same notation as those of Eq.(\ref{e14}), then we have the
following equation:

\begin{eqnarray}
{{d^2 f} \over {d \xi^2}}+f^2 {df \over {d \xi}}-f=0.  \label{e18}
\end{eqnarray}
\noindent
We cannot solve exactly in this case either, but we can see that
$f(\xi) \sim \sqrt{2(\xi-\xi_0)}$\ as $\xi \rightarrow \infty$.

\vspace{1cm}


\subsection{Classical solution without matter in pure $R^2$ }

\vspace{0.5cm}

\indent

In this case we put $\displaystyle{ H(t)=f^2(t)}$,
then equation of motion to be solved becomes in the form
$\displaystyle{ {{d^2 f} \over {dt^2}} +{df^3 \over dt} }=0$, in addition
to the trivial case $f=0$.  The first integral of this equation gives
$\displaystyle{ {df \over dt} +f^3 =-c_1^3}$, where $c_1$ is the integration
constant.  The special solution of this equation is given by
$f=-c_1=({\rm const.})$, which gives the de Sitter universe
$a(t)=a(0)\exp{ {c_1}^2 t }$.  Most general solution is given in the
following implicit form:

\begin{eqnarray}
{1 \over {6 c_1^2}} \log{\left({ {(f+c_1)^3} \over {f^3+c_1^3}}\right)}
+{1 \over {\sqrt{3} c_1^2}} \arctan{{2f-c_1} \over {\sqrt{3} c_1}}=(t_0-t).
\label{e19}
\end{eqnarray}

In $c_1=0$ case, we can express $f$\ in the explicit form
${ f={1 / \sqrt{2(t-t_0)}} }$, which gives
the square root type solution of the universe

\begin{eqnarray}
a(t)=a_0 \sqrt{t-t_0}.
\label{e17-1}
\end{eqnarray}

\vspace{1cm}


\section{Another Formulation of $R^2$ Cosmology}

\vspace{0.5cm}

\indent

In order to reduce the higher derivative term, we introduce an auxiliary
field $\sigma$ and rewrite the original action in the form:

\begin{eqnarray}
I&~=\displaystyle{ \int} d^4x \sqrt{-g} \left[
-\displaystyle{ \frac{1}{16 \pi G}}
\left(R + \alpha (- 2R \sigma-{\sigma}^2 ) \right)
+{\cal L_{\rm matter}} \right]~. \label{e20}
\end{eqnarray}

This reproduces the original action after substituting the solution of
$\sigma$\ field.  In the quantum theory, taking the path
integral formulation, we have additional term
$\prod_{x} 1/\left[-g(x)\right]^{1/4}$\ when we integrate over $\sigma$.
This additional factor changes the measure of the inner product, which
is connected to way how to define the ordering of operators.
In this sense, our quantum theory of $R^2$\ gravity and conventional quantum
 theory of $R^2$\ is different only in the measure.  As there is no
{\it a priori} criterion to determine the measure, here we take above
action as the starting action in quantum gravity.

Using the metric of Eq.(\ref{e2}), we obtain

\begin{eqnarray}
I=\displaystyle{\int} d^3 x \displaystyle{\int} dt
 \left[ - {1 \over {16\pi G}}
     \left\{  6{{a \dot{a}^2}  \over  b}
 +\alpha a^3 b \left(12\sigma \left( { \ddot{a} \over {a b^2}}
             +{ \dot{a}^2   \over {a^2 b^2}}
             -{ {\dot{a}\dot{b}}  \over  {a b^3}} \right)-\sigma^2
         \right)
    \right\}
    + a^3 b {\cal L_{\rm matter}}
 \right].
\label{e21}
\end{eqnarray}

In pure $R^2$ cosmology, we can analize exactly.  Then we will consider
only pure $R^2$ case hereafter. In pure $R^2$ cosmology, the Lagrangian
density becomes

\begin{eqnarray}
{\cal L}={{\alpha a^3 b} \over {16\pi G}}
     \left(-12\sigma \left( { \ddot{a} \over {a b^2}}
             +{ \dot{a}^2   \over {a^2 b^2}}
             -{ {\dot{a}\dot{b}}  \over  {a b^3}} \right)+\sigma^2
         \right).
    \label{e22}
\end{eqnarray}

We define $\displaystyle{ X_{+}=a^{3/2}\sigma^{3/2},
X_{-}=a^{3/2}, B=b\sigma^{1/2} }$, and after partial integral,
we obtain\cite{Takasugi}

\begin{eqnarray}
{\cal L}={\alpha \over {16\pi G}}
 \left\{{16 \over {3 B}}
          \dot{X_{+}} \dot{X_{-}}
          +B X_{+} X_{-}
\right\}.
    \label{e23}
\end{eqnarray}

Using this form of Lagrangian, we obtain equations of motion

\begin{eqnarray}
&&\dot{X_{+}} \dot{X_{-}}={{3 B^2} \over 16} X_{+} X_{-}, \nonumber \\
&&\ddot{X_{\pm}}={{3 B^2} \over 16} X_{\pm} .
    \label{e24}
\end{eqnarray}

Combining these equations, we can obtain general classical solutions in
pure $R^2$ cosmology by the following two steps.

\vspace{0.7cm}

\noindent
{\it Step 1)}
\noindent

First we use temporal time variable $T$, which is defined by the condition
$B(T)=1$, and solve equations of motion with this $T$. The metric at this
step gives

\begin{equation}
ds^2=-b(T)^2 {dT}^2+a(T)^2\left[ {dr}^2
+r^2 {d\theta}^2 +\sin{\theta}^2 {d\varphi}^2 \right]~.  \label{e25}
\end{equation}

\vspace{0.3cm}

\noindent
{\it Step 2)}
\noindent

The original time variable $t$ and the temporal time variable $T$ is
connected with the condition $dt=b(T)dT$, then the classical solution
of the original metric is given by $a(t)=a(T(t))$.

\vspace{0.7cm}

In Step 1, equations we deal with are followings:

\begin{eqnarray}
&&{{d X_{+}} \over {dT}} {{d X_{-}}  \over {dT}}=
{3  \over 16}  X_{+} X_{-}, \nonumber \\
&&{{d^2 X_{\pm}} \over {dT^2}}={3  \over 16} X_{\pm} ,
    \label{e26}
\end{eqnarray}

\noindent
which come from Eq.(\ref{e24}) after substituting $B=1$.
Then we get general solutions of the following three types:

\begin{eqnarray}
&&\hskip -1cm{\it{solution \ 1}}           \nonumber \\
&&X_{+}=X_{+}(0) \exp{\omega T} , \nonumber \\
&&X_{-}=X_{-}(0) \exp{\omega T} , \label{e27a} \\
&&\hskip -1cm{\it{solution \ 2}}           \nonumber \\
&&X_{+}=X_{+}(0) \cosh{\omega(T-T_0)} , \nonumber \\
&&X_{-}=X_{-}(0) \sinh{\omega(T-T_0)} , \label{e27} \\
&&\hskip -1cm{\it{solution \ 3}}           \nonumber \\
&&X_{+}=X_{+}(0) \sinh{\omega(T-T_0)} , \nonumber \\
&&X_{-}=X_{-}(0) \cosh{\omega(T-T_0)} , \label{e28}
\end{eqnarray}

\noindent
with $\omega=\sqrt{3/16}$. From these quantities, we construct
$a(T)=(X_{-})^{2/3}, b(T)=(X_{-}/X_{+})^{1/3}$.

In Step 2, solutions is  expressed with the original
time variable $t$ via the temporal time $T$ as follows:

\begin{eqnarray}
&&\hskip -1cm{\it{solution \ 1}}           \nonumber \\
&&t-t_0=(X_{-}(0)/X_{+}(0))^{1/3} T ,
\nonumber \\
&&a(T)=(X_{-}(0))^{2/3} \exp{2 \omega T /3}, \label{e29a} \\
&&\hskip -1cm{\it{solution \ 2}}           \nonumber \\
&&t-t_0=(X_{-}(0)/X_{+}(0))^{1/3} \int dT (\tanh{\omega (T-T_0)})^{1/3} ,
\nonumber \\
&&a(T)=(X_{-}(0) \sinh{\omega(T-T_0)})^{2/3} , \label{e29} \\
&&\hskip -1cm{\it{solution \ 3}}           \nonumber \\
&&t-t_0=(X_{-}(0)/X_{+}(0))^{1/3} \int dT (\coth{\omega (T-T_0)})^{1/3} ,
\nonumber \\
&&a(T)=(X_{-}(0) \cosh{\omega(T-T_0)})^{2/3} . \label{e30}
\end{eqnarray}

\noindent
In the above solutions, $X_{\pm}(0) \rightarrow
{\rm (finite)} \times \exp{\omega T_0}$\ with $T_0 \rightarrow -\infty$\
reproduce the previous solution
$\displaystyle{ a(t)=a_0 \sqrt{t-t_0} }$\ given in Eq.(\ref{e17-1}).

General solutions in these forms are more useful than the expression
in Eq.(\ref{e19}). This is because $f(t)$ is not given in the explicite
form in Eq.(\ref{e19}), and also because we must
integrate further to obtain $a(t)$ from
$\displaystyle{ f^2(t)={{da(t)/dt} \over a(t)} }$\ even if we know
the explicite form of $f$.

\vspace{1cm}


\section{Quantum solution in $R^2$~ cosmology}

\vspace{0.5cm}

\indent

When we solve classical solution in pure $R^2$\ gravity, we noticed that
the auxiliary field formalism is quite powerful.
Therefore, even in the quantum theory, we expect that it will become
easy to solve the Wheeler-DeWitt equation when we use auxiliary field
formalism.

Using the auxiliary field formalism, we start from
the previous Lagrangian Eq.(\ref{e23})

\begin{eqnarray}
{\cal L}=\hat{\alpha} \left\{{16 \over {3 B}}
          \dot{X_{+}} \dot{X_{-}}
          +B X_{+} X_{-}
\right\}
    \label{e31}
\end{eqnarray}

\noindent
with $\hat{\alpha}=\alpha/{16\pi G}$. We define the canonical momentum
in a standard way

\begin{eqnarray}
&&P_{+}={ {\partial {\cal L}} \over {\partial \dot{X_{+}}} }=
{ {16 \hat{\alpha}} \over {3B}  } \dot{X_{-}}, \nonumber \\
&&P_{-}={ {\partial {\cal L}} \over {\partial \dot{X_{-}}} }=
{ {16 \hat{\alpha}} \over {3B}  } \dot{X_{+}}, \nonumber \\
&&P_{B}={ {\partial {\cal L}} \over {\partial \dot{B}} } \approx 0 \quad
(\rm primary \ \ constraint).  \label{e32}
\end{eqnarray}

The Hamiltonian is given as follows:

\begin{eqnarray}
{\cal H}=P_{+}\dot{X_{+}}+P_{-}\dot{X_{-}}-{\cal L}
= {3B  \over {16 \hat{\alpha}}} P_{+} P_{-} -\hat{\alpha} B X_{+} X_{-}
\label{e33}
\end{eqnarray}

The secondary constraint, which comes from $\dot{P_{B}} \approx 0$,
gives

\begin{eqnarray}
 \dot{P_{B}}=\{{\cal H}, P_{b} \}_{\rm cl.}
={3  \over {16 \hat{\alpha}}} P_{+} P_{-} -\hat{\alpha}
X_{+} X_{-} \approx 0 .
\label{e34}
\end{eqnarray}

Taking the Schr\"{o}dinger representation $\displaystyle{
P_{+}=-i {\partial \over { \partial
X_{+}} },\quad P_{-}=-i {\partial \over { \partial X_{-}} }  }$,
Eq.({\ref{e34}) gives the Wheeler-DeWitt equation of the form

\begin{eqnarray}
\left[    {\partial \over { \partial X_{+}} }
          {\partial \over { \partial X_{-}} }
+{{16 \hat{\alpha}^2} \over 3} X_{+} X_{-} \right] \Psi(X_{+},X_{-})=0 .
\label{e36}
\end{eqnarray}

In the above equation, the operator in square brackets is symmetry
under the transformation

\begin{eqnarray}
&&X_{+} \rightarrow \ell X_{+}, \nonumber \\
&&X_{-} \rightarrow {1 \over \ell} X_{-}, \label{e36-1}
\end{eqnarray}

\noindent
where $\ell$ is an arbitrary constant.
If we find a solution $\Psi(X_{+},X_{-})$,
$\displaystyle{\Psi(\ell X_{+},{1\over\ell}X_{-})}$ also becomes
a solution by using the transformation of Eq.({\ref{e36-1})
in Eq.(\ref{e36}).

It is convenient to rewrite with variables $X_0,\ X_1$, which are defined by
$X_{+}={1\over \ell}(X_0 + X_1)$, $X_{-}=\ell(X_0 - X_1)$.
Using these variables, Eq.(\ref{e36}) becomes in the form

\begin{eqnarray}
\left[ \left( {\partial^2 \over { \partial X_{0}^2} }-
          {\partial^2 \over { \partial X_{1}}^2 } \right)
+{{64 \hat{\alpha}^2} \over 3} \left( X_{0}^2- X_{1}^2 \right)
\right] \Psi(X_{+},X_{-})=0 .
\label{e37}
\end{eqnarray}

In order to find the eigenstate of the state vector, we put the state
vector into the separable form
$\Psi(X_{+},X_{-})=\Psi^{0}(X_{0})\Psi^1(X_1)$. Using this separable form,
the above equation becomes separated into two Wheeler-DeWitt equations,
which are nothing but the Schr\"{o}dinger equation of the harmonic oscillator
with upside-down potential\cite{Guth}.

\begin{eqnarray}
\left( -{\partial^2 \over { \partial X_{0}^2} }
-{{64 \hat{\alpha}^2} \over 3}  X_{0}^2 \right) \Psi^0(X_0)=E \Psi^0(X_0)
\nonumber \\
\left( -{\partial^2 \over { \partial X_{1}^2} }
-{{64 \hat{\alpha}^2} \over 3}  X_{1}^2 \right) \Psi^1(X_1)=E \Psi^1(X_1)
\label{e38}
\end{eqnarray}

Solutions are given by

\begin{eqnarray}
\Psi^{0}(X_{0})=\Phi_{n}(X_{0}),\qquad \Psi^{1}(X_1)=\Phi_{n}(X_{1})
\label{e39}
\end{eqnarray}

\noindent
with

\begin{eqnarray}
&&\Phi_{n}(X)=H_{n}(e^{-i \pi/4} \beta X) e^{i \beta^2 X^2/2}, \nonumber \\
&&E=E_{n}=-2i \beta^2(n+1/2),\quad
\beta={\left({{64 \hat{\alpha}^2} \over 3}\right)}^{1/4}
\label{e39-1}
\end{eqnarray}

\noindent
These $\Phi_{n}(X)$\ are nothing but analytic continued forms
from those of the harmonic oscillator.
As equations are linear, superposition are also solutions. General
solutions are given in the form

\begin{eqnarray}
\Psi(X_{+},X_{-})=\sum^{\infty}_{n=0} \left[ c_n \Phi_{n}(X_0) \Phi_n(X_1)
+d_n \Phi_{n}^{*}(X_0) \Phi_{n}^{*}(X_1) \right]. \label{e39h}
\end{eqnarray}

Here we must comment on these solutions.  First of all, these state
vectors are not normalizable. The origin of this is because of the
instability of the gravitational theory.  According to the standard Copenhagen
interpretation of the quantum theory, the above solutions are unphysical.
But the above solutions are realy quite physical when we compare the above
quantum solutions with classical solutions, as we will discuss on that later.
Therefore, here in this paper, we take milder standpoint than
the standard Copenhagen interpretation especially in quantum cosmology,
and allow all solutions of the state vector without imposing
the normalizability.
As we do not impose the normalizability on the state vector, we obtain
twice as many solutions compared to those of the harmonic oscillator.

\vspace{1cm}


\subsection{Correspondence between classical and quantum solutions
under de Broglie-Bohm interpretation}

\vspace{0.5cm}

\indent

In this section, we study the connection between classical and quantum
solutions to understand the problem of time in the general relativity.
In the general relativity, because of the invariance under
general coordinate transformation, the consequent Hamiltonian becomes
weakly zero, then we come to the problem of time in canonical formalism.
This is the problem because we may naively think that the weakly zero
Hamiltonian will give only the solution of unrealistic static universe.

To overcome this unrealistic conclusion, many people take an approach from
the WKB approximations\cite{Banks,Alwis,Ohkuwa}.
The essential idea of the WKB approach is the followings.
We first separate the Hamiltonian into classical and quantum parts,
and we introduce time through the classical solution. Then the quantum
Hamiltonian plays the role of the time development operator in the
quantum system.  Though classical and quantum Hamiltonians are both
time developing, total Hamiltonian balanced to becomes zero.
If there is matter field in the system, this approach may be reasonable,
as the classical Hamiltonian may be dominated from the metric contribution
and the quantum Hamiltonian may be dominated from the matter field.
We cannot see the time dependence of the metric on the co-moving system,
the matter may develop with time according to the quantum Hamiltonian.

We will not take this standard approach in our pure $R^2$ cosmology. Because
there is no matter, it is unrealistic that classical and quantum contributions
from the metric have the same magnitude.  In the following, we take the
de Broglie-Bohm interpretation\cite{dBB}
between classical and quantum solutions and try to study the problem of
time.

\vspace{0.7cm}

In order to compare classical and quantum solutions in $R^2$\ cosmology,
we first consider the simplest state $\Psi_0$.
The state vector in this case is given by
$\displaystyle{ \Psi_0=e^{i \beta^2 (X_0^2+X_1^2)/2}=
e^{i \beta^2 (X_{+}^2+X_{-}^2)/4 } }$.
 In general, we decompose
the state vector into radial and phase components in the form
$\Psi=A e^{i S }$. Then $S$\ for $n=0$\ state is given by
$\displaystyle{S={{\beta^2}\over 4}(\ell^2 X_{+}^2+{1\over{\ell^2}}X_{-}^2)}$.

According to the de Bloglie-Bohm interpretation,
we obtain the first order differential equations for $X_{\pm}$ as

\begin{eqnarray}
&&{ {16 \hat{\alpha}} \over {3}  } \dot{X_{-}}=P_{+}=
 {{\partial S} \over {\partial X_{+}}}={{\ell^2 \beta^2 X_{+}} \over 2},
\nonumber \\
&&{ {16 \hat{\alpha}} \over {3}  } \dot{X_{+}}=P_{-}=
 {{\partial S} \over {\partial X_{-}}}={{\beta^2 X_{-}} \over {2\ell^2}}.
\label{e40}
\end{eqnarray}

\noindent
{}From the above equations and noticing relations
$\beta^2=8 \hat{\alpha} /\sqrt{3}$, $\omega=\sqrt{3/16}$, we obtain
$\dot{X_{+}}=\ell^{-2}\omega X_{-}$, $\dot{X_{-}}=\ell^{2}\omega X_{+}$.
Then we have the second order differential equation
$\ddot{X_{\pm}}=\omega^2 X_{\pm}$.
The general solution is given by

\begin{eqnarray}
X_{+}&=&{1\over \ell}(c_1 e^{\omega T} + c_2 e^{-\omega T}) \nonumber \\
X_{-}&=&{{\ell^2 \dot{X_{+}}} \over \omega}
=\ell(c_1 e^{\omega T} - c_2 e^{-\omega T}).
\label{e41}
\end{eqnarray}

If we compare these solutions with those of Eq.(\ref{e26}),
the above solutions correspond to the classical solutions,
Eqs.(\ref{e27a})-(\ref{e28}),exactly.  The more detailed correspondence
between the parameters in Eq.(\ref{e41}) and Eqs.(\ref{e27a})-(\ref{e28})
is given as follows with $\ell=\sqrt{ X_{-}(0)/X_{+}(0)}$:

\begin{eqnarray}
 &&\hskip -1cm{\it {solution \ 1}}(c_1>0, c_2=0) \nonumber \\
 && c_1=\sqrt{X_{+}(0) X_{-}(0)},
\nonumber \\
 &&\hskip -1cm{\it {solution \ 2}}(c_1>0, c_2>0) \nonumber \\
 && c_1=\sqrt{X_{+}(0) X_{-}(0)} e^{-\omega T_0},~
    c_2=\sqrt{X_{+}(0) X_{-}(0)} e^{\omega T_0},~
    \nonumber \\
 &&\hskip -1cm{\it {solution \ 3}} (c_1>0, c_2<0) \nonumber \\
 && c_1=\sqrt{X_{+}(0) X_{-}(0)} e^{-\omega T_0},~
    c_2=-\sqrt{X_{+}(0) X_{-}(0)} e^{\omega T_0}.~
     \label{e41-1}
\end{eqnarray}

Because of instability in $R^2$\ cosmology, that is, the potential of the
harmonic oscillator being
upside-down, the solution becomes the sum of
exponentially blowing up and damping term as time developes.
This behavior of the classical
solution is quite physical, we relax the standard Copenhagen
interpretation and allow the unrenormalizable state vector
in the quantum cosmology.

For general state vector $\Psi_{n}$ with $n \ge 2$, expressions obtained from
the de Broglie-Bohm interpretation is different from expressions of
classical solutions.  This is because only $n=0$\ and $n=1$\ state vecotor is
composed of only one term.

If the arguments of the Hermite polynomials become large,
which describes that the universe becomes large, expressions constructed
from the de Broglie-Bohm interpretation approach classical solutions.
This is because the additional phase of wave functions takes constant value
$- n \pi/ 2$ in the above case, and constant phase
does not contribute in $\partial S/\partial X_{\pm}$.

According to de Broglie-Bohm interpretation, time comes through the relation
\break ${ \dot{X}=P={{\partial S} / {\partial X}} }$, so
that even if the Hamiltonian
becomes zero system develops according to time.

In the rest of this section, we will point out the relation
between de Broglie-Bohm interpretation and Ehrenfest theorem\cite{Greensite},
which describes classical orbits as expectation values in quantum mechanics.
For the one dimensional Hamiltonian
$\displaystyle{ H=-{1 \over {2m}}{\partial^2 \over {{\partial x}^2}}
+V(x) }$, the Ehrenfest theorem gives

\begin{eqnarray}
{{d<X>}  \over {dt}} ={d \over {dt}} \int dx \psi^{\ast}(x,t) x \psi(x,t)=
-{i \over {2m}} \int dx \left( \psi^{\ast} {{\partial \psi}
\over {\partial x}}
-{{\partial \psi^{\ast}} \over {\partial x}} \psi \right)
\label{e43}
\end{eqnarray}

\noindent
after partial integration.  While, according to the de Broglie-Bohm
interpretation, we obtain

\begin{eqnarray}
m{dX \over {dt}}=
\left.
-{i \over {2|\psi|^2}} \left( \psi^{\ast} {{\partial \psi} \over {\partial x}}
-{{\partial \psi^{\ast}} \over {\partial x}} \psi \right)
\right|_{x=X}=
\left.
{{\partial S} \over {\partial x}}
\right|_{x=X}.
\label{e44}
\end{eqnarray}

\noindent
If we compare Eq.(\ref{e43}) with Eq.(\ref{e44}), we can understand that
de Borglie-Bohm interpretation
is the {\it local} form of Ehrenfest theorem.
Here we emphasize that the state vector is assumed to be
normalizable in the proof of Ehrenfest theorem, while the normalizability of
the state vector is not assumed in the local form of Eq.(\ref{e44}).
In fact, the state vector is not normalizable in our examples.

\vspace{1cm}

\section{Summary}

\vspace{0.5cm}

\indent

In this paper, we have studied various classical solutions in the
generalized Einstein cosmology which contains higher derivative terms $R^2$.
Especially we obtained general classical solutions in pure
$R^2$\ cosmology.
Even in the quantum theory, we can solve the Wheeler-DeWitt equation
in pure $R^2$\ cosmology exactly.
We compared classical and quantum solutions in this $R^2$ cosmology,
and studied the problem of time using the de Broglie-Bohm interpretation.
Further we pointed out the relation between de Broglie-Bohm
interpretation and the Ehrenfest theorem.

\vspace{1.5cm}

\noindent
{\large \bf Acknowledgement}:
  Two of us (K.S. and K.U.) are grateful to the special research funds at
Tezukayama University.

\newpage


\end{document}